\documentclass[letterpaper,conference]{IEEEtran}
\usepackage{color}
\usepackage{epsfig}
\usepackage{amsmath}
\usepackage{amsfonts}
\usepackage{amsthm}
\usepackage{amssymb}
\usepackage{graphics}
\usepackage{graphicx}
\usepackage{float}
\usepackage{algorithm}
\usepackage{algorithmic}
\usepackage{svg}
\usepackage {epstopdf}
\usepackage{stfloats}
\usepackage{flushend}

\title{\LARGE \bf A modified probabilistic amplitude shaping scheme to use sign-bit-like shaping with a BICM}

\author{Vincent Corlay and Hamidou Dembele\\
\small{Mitsubishi Electric R\&D Centre Europe, Rennes, France. E-mail: v.corlay@fr.merce.mee.com.} }

\begin{document}

\maketitle

\begin{abstract}
On the one hand, sign-bit shaping is a popular shaping scheme where the conditional probability of the sign bit is made non-equiprobable.
On the other hand, probabilistic amplitude shaping (PAS) is a popular coding scheme, to combine shaping and a bit-interleaved coded modulation (BICM), where the sign bit should not be involved in the shaping. 
Indeed, with the PAS scheme the sign bit is the parity bit, i.e., the output of the systematic error-correcting code.
As a result, sign-bit shaping has been used with multilevel coded modulations rather than BICM. 
In this paper, we show that with minor modifications it is possible to use sign-bit-like shaping with a BICM.
Simulation results are provided with the 5G NR LDPC BICM scheme.
\end{abstract}

\begin{IEEEkeywords}
Probabilistic shaping, sign-bit shaping, bit-interleaved coded modulation.
\end{IEEEkeywords}


\section{Introduction}
\label{sec_intro}

The channel capacity characterizes the highest information rate that can be achieved for a fixed average transmit power while maintaining a small error probability.

When standard constellations are used, such as the amplitude-shift keying (ASK), the channel capacity cannot be reached if each symbol is transmitted with equal probability. 
Hence, the transmitter should process the data such that the symbols of the constellation are transmitted according to a probability distribution which enables to approach the capacity. 
This operation is called probabilistic shaping.

In addition to shaping, the message should also be protected with an error-correcting code. 
Combining shaping and coding is not trivial and requires a specific algorithm. 

There exist two main techniques to build high-rate coded modulations: the BICM \cite{Caire1988}\cite{Bocherer2015} and multilevel coding \cite{Imai1977}\cite{Wachsmann1999}\cite{Corlay2022B}. 
The popular PAS scheme \cite{Bocherer2015} (see Section~\ref{sec_PAS_scheme}), to combine shaping and coding, uses a BICM. 
With the PAS scheme, the parity bits, at the output of the error-correcting code, are used as sign bits (i.e., the bit determining the sign of the symbols).
Consequently, the sign bit cannot be considered for the shaping operation as its distribution is independent of the value of the other labelling bits of the symbol.

Nevertheless, ``sign-bit shaping" is a popular shaping technique where the conditional distribution of the sign bit is made non-equiprobable \cite{Forney1992}\cite{Wachsmann1999}\cite{Bohnke2020}\cite{Corlay2022}.
Hence, it cannot be considered as a distribution matcher (DM) for the PAS scheme. As a result, (to the best of our knowledge) sign-bit shaping has always been considered jointly with multilevel coding\footnote{Sign-bit shaping is conveniently implemented with multilevel coding as the last level (the sign-bit level with natural labelling) does not need to be coded. Indeed, the mutual information of the last bit level equals the entropy.}.
As an example, in our previous paper on sign-bit shaping \cite{Corlay2022}, a reviewer asked us to explicitly state that the proposed technique is restricted to a multilevel coding scheme.

In this paper, we propose modifications to the conventional PAS scheme (Section~\ref{sec_modif_PAS_scheme}) and to the probabilistic sign-bit shaping scheme proposed in \cite{Corlay2022} (Section~\ref{sec_modif_signbit}).
These modifications enable to use the paradigm of sign-bit shaping with a BICM.
We also describe a mechanism to make the shaping scheme compatible with the puncturing of systematic bits, as done e.g., in the current 5G NR LDPC BICM scheme (Section~\ref{sec_decrease_rate}).


\section{The PAS scheme}
\label{sec_PAS_scheme}

\begin{figure}
\centering
\includegraphics[width=1\columnwidth]{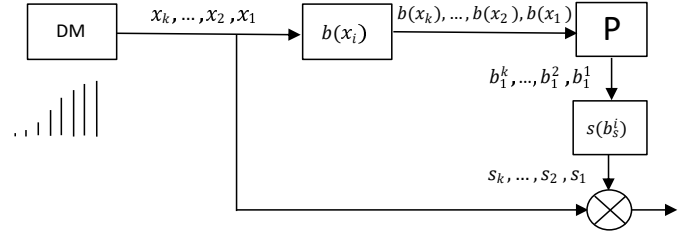}
\caption{PAS scheme.  The DM generates symbols according to one side of a target shaping distribution. The function $b(\cdot)$ outputs the labelling bits of a given symbol $x_i$. The block “P” computes and outputs the parity bits (i.e., implements the systematic error-correcting code). The function $s(\cdot)$ outputs the sign corresponding to a bit  $b_s^i$.}
\label{fig_Shaping_PAS}
\end{figure}

The principle of PAS, illustrated on Figure~\ref{fig_Shaping_PAS}, is the following: A DM outputs symbols according to one side (negative or positive) of the target shaping distribution.
Then, the bits corresponding to the labelling of the symbols\footnote{Each symbol is labelled with several bits. See Figure~\ref{fig_table_nat_lab} for an exemple, where bit level 4 is the sign bit.}, without the sign bit, are used as inputs of a systematic error-correcting code. 
The encoding process outputs parity bits, one per symbol, which determine the sign of the symbols to be transmitted. 
The key ideas underlying PAS are the following:
\begin{itemize}
\item	Since systematic encoding is used, the distribution of the shaping bits is not changed by the error-correcting code. They can be non-i.i.d.
\item	The parity bits of an error-correcting code have an equiprobable distribution \cite[Theorems 1,2]{Wonterghem2018}. This is suited to symmetric shaping distributions: The symbols have the same probability to be positive and negative and the sign bits should therefore remain equiprobable. 
\end{itemize}

Consequently, the PAS scheme successfully combines shaping and coding.

\begin{figure*}[t]
\begin{center}
\begin{tabular}{|c|c|c|c|c|c|c|c|c|c|c|c|c|c|c|c|c|}
\hline
Symbols	&-15 &	-13	&-11	&-9	&-7	&-5	&-3	&-1	&1	&3&	5&	7	&9	&11	&13	&15\\
\hline
Bit-level 1 ($b_1$)	& 0	&1	&1    &0	&0&	1&	1	&0&	0	&1	&1	&0&	0	&1&	1	&0\\
\hline
Bit level 2 	($b_2$) &0	&0	&1	&1	&1	&1	&0	&0	&0	&0	&1	&1	&1	&1	&0	&0\\
\hline
Bit level 3	($b_3$) &0	&0	&0	&0	&1	&1	&1	&1	&1	&1	&1	&1    &0	&0	&0	&0	 \\
\hline
Bit level 4	($b_4$) &0	&0	&0	&0	&0	&0	&0	&0	&1	&1	&1	&1	&1	&1	&1	&1  \\
\hline
\hline
\hline
Bit-level 1 ($b_1$)	& 0	&1	&0    &1	&0&	1&	0	&1&	0	&1	&0	&1&	0	&1&	0	&1\\
\hline
Bit level 2 	($b_2$) &0	&0	&1	&1	&0	&0	&1	&1	&0	&0	&1	&1	&0	&0	&1	&1\\
\hline
Bit level 3	($b_3$) &0	&0	&0	&0	&1	&1	&1	&1	&0	&0	&0	&0    &1	&1	&1	&1	 \\
\hline
Bit level 4	($b_4$) &0	&0	&0	&0	&0	&0	&0	&0	&1	&1	&1	&1	&1	&1	&1	&1  \\
\hline
\end{tabular}
\end{center}
\caption{Gray labelling (top) and natural labelling (bottom) of a 16-ASK.}
\label{fig_table_nat_lab}
\end{figure*}

\section{The modified PAS scheme}
\label{sec_modif_PAS_scheme}

\subsection{The quantized Maxwall-Boltzmann like distribution}
The constellation $\mathcal{X}$ considered in this paper is a $M$-ASK constellation. The symbols of a $M$-ASK constellation, where $M=2^m$, are
\begin{align}
\mathcal{X}=\{-2^m+1,..,-3,\ -1,+1,+3,\ldots,+2^m-1\}.
\end{align}

\begin{figure}[h]
\centering
\includegraphics[width=1\columnwidth]{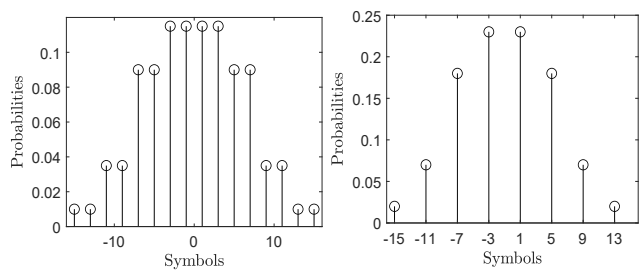}
\caption{Left: Example of a quantized target shaping distribution for $\mathcal{X}$. \\
Right: Target shaping distribution for the sub-constellation $\mathcal{X}_r$}
\label{fig_quantized_distri}
\end{figure}

The discrete Maxwell-Boltzmann (MB) distribution, which is a quasi-optimal input distribution for the symbols of $\mathcal{X}$ on the Gaussian channel \cite{Kschischang1993}, can be quantized at the cost of negligible performance loss \cite{Gultekin2019}\cite{Corlay2022}.  
As an example, the distribution of the 16-ASK constellation shown on Figure~\ref{fig_quantized_distri} (left) exhibits quasi-optimal performance in terms of mutual information:
The loss is less than 0.1 dB for information rates smaller than 3 bits per channel use (bpcu), see Figure~4 in \cite{Corlay2022}.

Consequently, this quantized MB-like distribution can be taken as the target shaping distribution.

\subsection{Using the quantification bit as parity bit}

\subsubsection{The principle}
The quantized constellation $\mathcal{X}$ can be expressed as the union of two shifted versions of a reference sub-constellation, say~$\mathcal{X}_r$.
\begin{align}
\mathcal{X} = \mathcal{X}_r\ \cup\ {(\mathcal{X}}_r\ +2).
\end{align}
With the 16-ASK, $\mathcal{X}_r=\{-15,-11,-7,-3,1,5,9,13\}$.
Moreover, a transmitted symbol belongs equiprobably to one of the two sub-constellations.
Consequently, one can proceed as follows:
\begin{itemize}
\item First, perform the shaping of the target sub-constellation $\mathcal{X}_r$, shown on Figure~\ref{fig_quantized_distri} (right).
\item Then, obtain the $m-1$ bits corresponding to each shaped symbol. Use them as input of a systematic error-correcting code.
\item Finally, use the parity bit as quantification bit, i.e., to decide if the symbol belongs to the first or the second sub-constellation.
\end{itemize}

Consequently, the major difference between the PAS scheme and the proposed scheme is the following: The parity bit discriminates between the two sub-constellations. It does not determine the sign of the symbols.

\subsubsection{Gray labelling}

If the coding scheme is a BICM, the labelling of the symbols has a significant impact on the performance.
For instance, Figure 8 in Sec VI.C of \cite{Bocherer2015} reports a 1 dB loss with natural labelling compared to Gray labelling.
We also observed this difference in our simulations.

Consequently, unlike in \cite{Corlay2022} where natural labelling (suited to a multilevel coding scheme) is used, Gray labelling of the symbols must be considered.
For $\mathcal{X}$ chosen as the $M=16$-ASK, this latter labelling is provided in Figure~\ref{fig_table_nat_lab} (top). Natural labelling is also shown on the figure (bottom).

As with natural labelling, given a symbol $x_i \in \mathcal{X}_r$ the bit $b_1$ discriminates between the two sub-constellations: All adjacent symbols with the same probability (according to Figure~\ref{fig_quantized_distri} (left)) have a different value for $b_1$. Consequently, the bits $b_2$,$b_3$,$b_4$ are used to label the symbols in $\mathcal{X}_r$.

\begin{figure}[h]
\centering
\includegraphics[width=1\columnwidth]{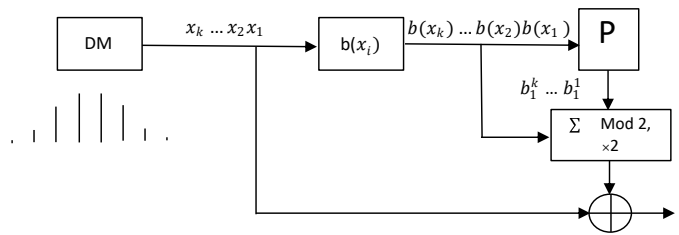}
\caption{Modified PAS scheme. The DM generates symbols according to the distribution of the sub-constellation $\mathcal{X}_r$. The function $b(\cdot)$ outputs the labelling bits of a given symbol $x_i$. The block “P” computes and outputs the parity bits.}
\label{fig_modif_Shaping_PAS}
\end{figure}

However, unlike with natural labelling, the rule to discriminate between the two sub-constellations depends on the value of $b_2$,$b_3$,$b_4$.
Given $x \in \mathcal{X}_r$:
\begin{itemize}
\item If $b_1 \oplus b_2 \oplus b_3 \oplus b_4= 0$ then transmit $x$.
\item If $b_1 \oplus b_2 \oplus b_3 \oplus b_4= 1$ then transmit $x+2$.
\end{itemize}
This modified PAS scheme is illustrated on Figure~\ref{fig_modif_Shaping_PAS}.

This trick enables for instance to use sign-bit shaping via trellis shaping with a BICM (described in \cite{Forney1992} and applied in the famous paper \cite{Wachsmann1999} with multi-level coding).
In the following section, we discuss the adaptation of another sign-bit shaping scheme: probabilistic sign-bit shaping as introduced in \cite{Corlay2022}.

Note also that this modified PAS scheme allows the shaping of (quantized) non-symmetric distributions, which is not possible with the PAS scheme.

\begin{figure*}
\centering
\includegraphics[width=1.6\columnwidth]{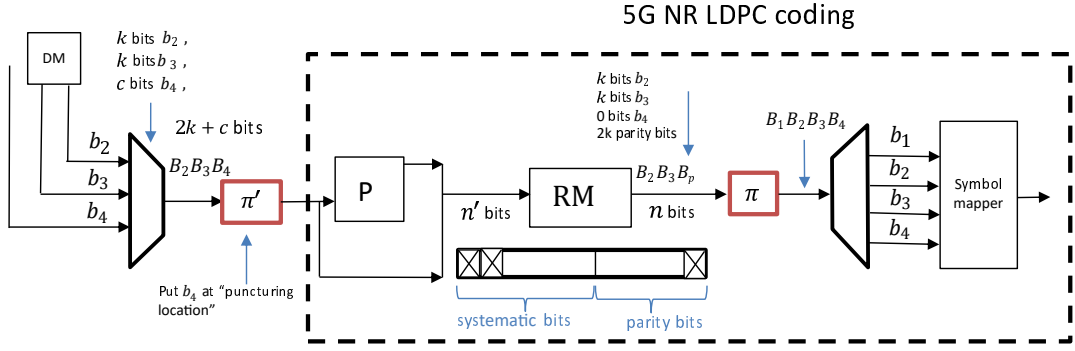}
\caption{Proposed scheme to decrease the rate of the binary code, for a 16-ASK. The right-most part is compliant with the 5G NR BICM scheme, where the 5G NR interleaver (row/column) is represented by a demultiplexer. 
The block “P” computes and outputs the parity bits (i.e., implements the systematic error-correcting code) and the puncturing is performed by the rate matcher (RM).
The only difference with the standard, in this right-most part, is the block $\pi$, which puts some parity bits at the location of the punctured systematic bits.
}
\label{fig_puntu_rate}
\end{figure*}

\subsection{Decreasing the code rate and puncturing systematic bits}
\label{sec_decrease_rate}

With the PAS scheme, the baseline rate of the code used is $R\ =\ \frac{m-1}{m}$.  Higher rates codes can be considered by using some sign bits as systematic bits (see Sec.~IV.D in \cite{Bocherer2015}). However, it is not possible to use lower rate codes with this standard scheme. Moreover, some BICM coding schemes include the puncturing of some systematic bits.
It is the case of the 5G NR LDPC coding scheme where the first systematic bits are always punctured (see e.g., Chap. 9 in \cite{Dahlman2018}).
We show below how the quantification bit enables to address both issues.

Let us consider a shaping scheme which does not change the distribution of the sign bit and the quantification bit (i.e., performs the shaping via $b_2$ and $b_3$ on Figure~\ref{fig_table_nat_lab}).  Moreover, let us assume that $c$ systematic bits are punctured by the coding scheme. Let $k$ be the number of symbols transmitted. Then, one can proceed as follows:
\begin{itemize}
\item Use $c \leq k$ sign bits and/or quantification bits as systematic bits (in addition to the other bits).
\item Put these bits at the systematic puncturing location.
\item Generate $2k$ parity bits (with the channel code).
\item Puncture $c$ systematic bits.
\item Put the parity bits at the proper location (the one of $b_1$ and $b_4$ on Figure~\ref{fig_puntu_rate}).
\end{itemize}
Figure~\ref{fig_puntu_rate} summarizes the process in the scope of the 5G NR LDPC coding scheme. The rate of the code is $R= \frac{(m-1-q)k +c}{mk+c} \leq \frac{m-1}{m}$, where $q \geq 1$ is the number of quantification bits.

Note that one could also have $c'$ bits $b_4$, $c< c' \leq k$, instead of the exact number of systematic punctured bits $c$, and diminish the number of generated parity bits accordingly.

\begin{figure}
\centering
\includegraphics[width=0.95\columnwidth]{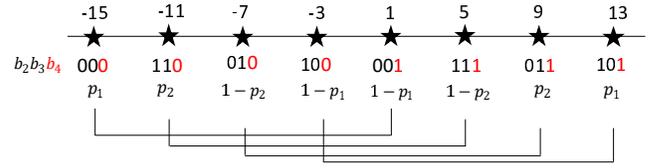}
\caption{Sub-constellation $\mathcal{X}_r$ with natural labelling and corresponding values~$p_i$. 
}
\label{fig_lab_ex}
\end{figure}

\begin{figure}
\centering
\includegraphics[width=0.95\columnwidth]{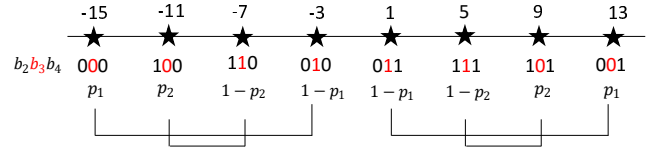}
\caption{Sub-constellation $\mathcal{X}_r$ with Gray labelling and corresponding values~$p_i$. 
}
\label{fig_lab_ex_bis}
\end{figure}

\section{Modified probabilistic sign-bit shaping}
 \label{sec_modif_signbit}

We now discuss a specific implementation of the DM of Figure~\ref{fig_modif_Shaping_PAS} to implement the target distribution of $\mathcal{X}_r$. 

With sign-bit shaping\footnote{See \cite{Corlay2022} for a more detailed explanation of sign-bit shaping.}, the probability of the sign bit depends on the values of the bits at the previous levels.
Sign-bit shaping for the distribution of $\mathcal{X}_r$ can be realized as follows. 

As mentioned above, since this sub-constellation has only 8 symbols (or more generally $M'=2^{m'}=M/2$ symbols, with $m'=m-1$), only $b_2$,$b_3$,$b_4$ are used for the labelling. Let us first consider the natural labelling, shown in Figure~\ref{fig_table_nat_lab} (bottom), as in \cite{Corlay2022}. 
Then, sign-bit shaping consists in adapting $p(b_4|b_1,b_2)$. The probability of each symbol of $x_i \in \mathcal{X}_r$ becomes
\small
\begin{align}
\begin{split}
&\forall \ 1 \leq i \leq \frac{M'}{2}, \  p(x_{i}) = p_i \cdot \left(\frac{1}{2} \right)^{m'-1}, \\
&\forall \ \frac{M'}{2} + 1 \leq i \leq M', \ p(x_{i}) = (1 - p_{i-\frac{M'}{2}}) \cdot \left(\frac{1}{2} \right)^{m'-1},
\end{split}
\end{align}
\normalsize
where the parameters $p_i = p(b_4|b_2,b_3)$, $0\leq p_i \leq 1$, are to be optimized. 
An illustration is provided with $\mathcal{X}_r$ on Figure~\ref{fig_lab_ex}. 
Since the target shaping distribution is symmetric, $p_3$ and $p_4$ are replaced by $1-p_2$ and $1-p_1$ on the figure. The distributions of Figure~\ref{fig_quantized_distri} are obtained with $p_1=0.08$ and $p_2=0.28$.

Unfortunately, with Gray labelling this mapping does not hold. For instance, the two symbols of $\mathcal{X}_r$ with $b_2=0,b_3=0$ are -15 and 13, which should have the same probability.
Nevertheless, if we replace $p_i=p(b_4|b_2,b_3)$ by $p_i=p(b_3|b_2,b_4)$  (or simply permute bit level 3 and bit level 4 in the Gray labelling) we get the mapping of Figure~\ref{fig_lab_ex_bis} which does not change the distribution of the symbols. As a result, we obtain a sign-bit-like shaping scheme where the probability of bit level 3 of the Gray labelling, conditioned on the value of $b_2$ and $b_4$, is non-equiprobable.

With this new mapping, we see that $p(b_3|b_2,b_4)=p(b_3|b_2)$. Hence, the sign bit $b_4$ can be removed from the shaping process. 
Note that this would not be the case if the target shaping distribution was not symmetric.


\begin{figure*}
\centering
\includegraphics[width=1.4\columnwidth]{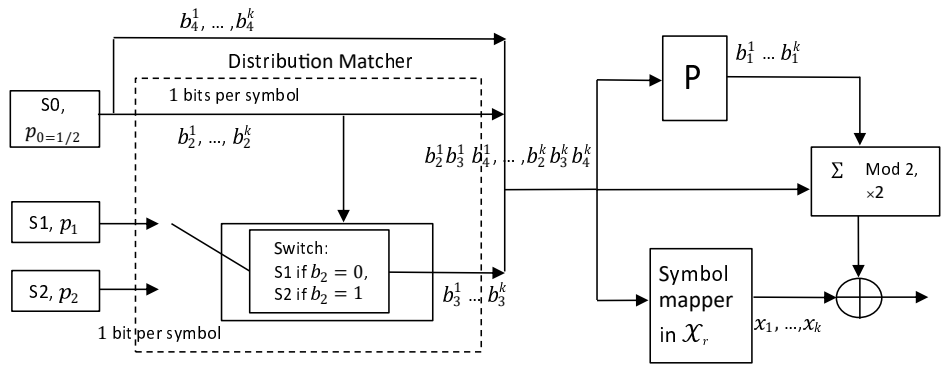}
\caption{Modified PAS scheme with sign-bit-like shaping. The sources $S_0$, $S_1$, and $S_2$ generate a bit equal to 0 with probability $1/2$, $p_1$, and $p_2$, respectively. The non-equiprobable sources $S_1$ and $S_2$ can be obtained from the binary source $S_0$ via binary DMs (see also \cite{Corlay2022}).}
\label{fig_modif_PAS}
\end{figure*}

\subsection{Implementation}

Regarding the implementation of the ``sign-bit-like shaping" DM, we proceed as follows:
The second and fourth bit levels are equiprobable and independent. 
Therefore, a binary source $S_0$ outputs two bits with equiprobable probability. 
The binary source for $b_3$ is chosen based on the value of $b_2$ (and is independent of the value of $b_1$ and $b_4$), i.e., $S_1$ has a distribution $p_1 = p(b_3|b_2=0)$ and $S_2$ a distribution $p_2 = p(b_3|b_2=1)$. 
Then, a symbol mapper outputs the symbols $x_i \in \mathcal{X}_r$ based on the values of $b_2^i, b_3^i, b_4^i$ and according to the Gray labelling.
Finally, $x_i$ is shifted if $b_1^i \oplus b_2^i \oplus b_3^i \oplus b_4^i =1$.
The full system is shown on Figure~\ref{fig_modif_PAS} (with $c'=k$).

\section{Simulation results}

For the simulations, we use the 5G NR systematic LDPC code \cite{Urbanke}.
The baseline rate of the code is 1/3. 
As mentioned above, this rate is obtained by (always) puncturing the first $2 Z$ systematic bits, where $Z$ is the lifting value which depends on the block length used \cite{3GPP_2}. 
Hence, the standard PAS scheme cannot be used (without altering the distribution of the symbols) with the 5G NR LDPC coding scheme. We need the trick of Section~\ref{sec_decrease_rate}.\\
Rate Matching, to increase the rate, is done as specified in TS 38.212 \cite{3GPP_2} by discarding the last parity bits.
As reported e.g., in  R1-1706971 (by Huawei) \cite{3GPP_1} (and confirmed by our own simulations) for a block length $n=7875$ bits and a rate $R=0.75$ (non-BICM case), the considered code achieves a block error rate of 10$^{-2}$ at a SNR of approximately 4 dB (1.4 dB away from the Shannon limit).

%

For the implementation of the shaping, we puncture the $2Z$ first systematic bits and keep $k+2Z$ parity bits after the RM. The extra $2Z$ parity bits are used as the missing sign bits $b_4$.

\begin{figure}
    \centering
    \includegraphics[width=1\columnwidth]{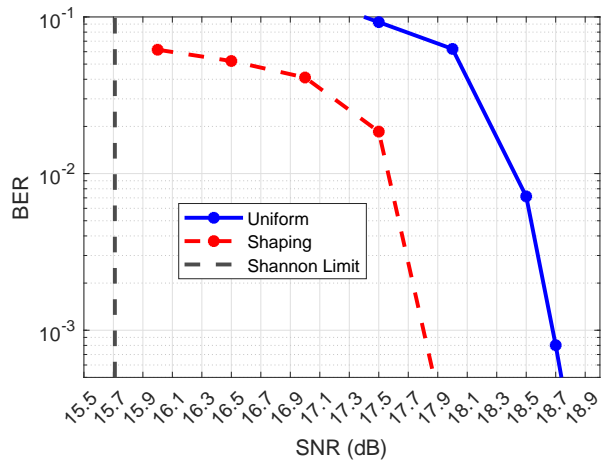}
    \caption{Performance of the 5G NR LDPC BICM scheme with and without shaping for a rate $R=2.63$ bpcu and block length $n=7875$ bits.}
    \label{BERPerformance5GStandard16ASK}
\end{figure}

Figure~\ref{BERPerformance5GStandard16ASK} presents the simulation results for a rate $R=2.63$ bpcu with a 16-ASK. The blue curve shows the performance of the standard 5G NR LDPC BICM scheme. The red curve shows the performance of the same system with shaping as described in the paper. We observe a gain of approximately 0.9 dB. This is consistent with what is expected: The information-theoretic study (see Figure~4 in \cite{Corlay2022}) tells us that the optimal shaping gain at this rate is 1 dB and 0.1 dB is lost due to the quantified shaping distribution.

%


\section{Conclusions}
In this paper, we showed how the quantified target distribution adds flexibility to the PAS scheme:
1-It enables to use the quantification bit as parity bit and thus use the sign bit as shaping bit if needed. 
2-It can be used to decrease the rate of the code. 
3-It also allows to have an i.i.d systematic bit, useful in the case of systematic bit puncturing (as e.g., in the 5G NR coding scheme). 
Moreover, we explained how probabilistic sign-bit shaping, originally used with natural labelling and thus multilevel coding, can be adapted to Gray labelling suited to a BICM.
Finally, simulation results are provided with the 5G NR LDPC BICM scheme.

\end{document}